\documentclass[aps,prl,amsmath,floats,floatfix,superscriptaddress,nofootinbib,noshowpacs,10pt,twocolumn]{revtex4-2}
\usepackage{amssymb,epsfig}
\usepackage{tensor}

\begin{document}

\title{Bondi-type accretion onto a Kerr black hole in the kinetic regime}
\author{Patryk Mach}
\email{patryk.mach@uj.edu.pl}
\affiliation{Instytut Fizyki Teoretycznej, Uniwersytet Jagiello\'{n}ski, {\L}ojasiewicza 11, 30-348 Krak\'{o}w, Poland}

\author{Mehrab Momennia}
\email{momennia1988@gmail.com}
\affiliation{Instituto de F\'isica y Matem\'aticas, Universidad Michoacana de San
Nicol\'as de Hidalgo, Edificio C-3, Ciudad Universitaria, 58040 Morelia,
Michoac\'an, M\'exico}

\author{Olivier Sarbach}
\email{olivier.sarbach@umich.mx}
\affiliation{Instituto de F\'isica y Matem\'aticas, Universidad Michoacana de San
Nicol\'as de Hidalgo, Edificio C-3, Ciudad Universitaria, 58040 Morelia,
Michoac\'an, M\'exico}

\begin{abstract}
We derive an exact solution representing a Bondi-type stationary accretion of a kinetic (Vlasov) gas onto the Kerr black hole. The solution is exact in the sense that relevant physical quantities, such as the particle current density or the accretion rates, are expressed as explicit integrals, which can be evaluated numerically. We provide an analytic approximation which allows us to obtain simple formulas for the mass, energy, and angular momentum accretion rates. These formulas are used to derive characteristic time scales of the black hole mass growth and the associated spin-down in two different scenarios: assuming that the ambient energy density is either constant or decreases on a cosmological scale.
\end{abstract}

\maketitle


\textit{Introduction.} The search for an exact solution representing a Bondi-type accretion in Kerr spacetime is a long-standing challenge. In 1988 Petrich, Shapiro, and Teukolsky obtained an analytic solution, assuming a zero-vorticity potential flow and the ultra-hard equation of state \cite{lPsSsT1988}. Their method was based on the observation that for the ultra-hard equation of state, the equations of motion governing zero-vorticity perfect fluids can be expressed in terms of a potential, and they are linear. More precisely, the continuity equation can be written as the massless Klein-Gordon equation, which is separable on the Kerr background spacetime. Unfortunately, no straightforward generalizations to perfect fluids with more realistic equations of state seem to be possible (see \cite{vByP95,vP96,eTpTjM13} for attempts to obtain approximate solutions).

In this letter, we present a new exact solution obtained for a gas described by the collisionless Boltzmann (Vlasov) equation, which is also linear. For sufficiently symmetric flows, it possesses simple solutions with the one-particle distribution function expressed in terms of constants of the geodesic motion. The difficulty in obtaining physically relevant results is related to controlling the regions of the phase space that are available for the motion of gas particles originating from the asymptotic region.

By a Bondi-type accretion we mean a nearly spherically symmetric, stationary flow of gas onto the black hole. The gas extends to infinity, where it is assumed to be homogeneous and at rest. Within the general-relativistic kinetic approximation, such models were investigated in \cite{pRoS16,pRoS17,aCpM20,BardeenBH,SchwLikeBH,KalbRamondBH,mMoS25}, assuming spherical symmetry. Properties of other kinetic models in the Kerr spacetime were analyzed in \cite{pRoS18,aCpMaO22,pRoS24,gKpM25,mLyYaC25}. Numerical simulations of accretion flows of the collisionless magnetized gas onto Kerr black holes were performed in \cite{kPaPbC19,aBbRaP21,bCbCgDkPaP22, aGaPeQfBkPbR23}.

In what follows, we discuss the main aspects of our analysis. Mathematical details and further results can be found in the accompanying paper~\cite{mainpaper}. We use geometrized units in which the speed of light and the gravitational constant are set to one.\\


\textit{Kinetic description of matter in the Kerr spacetime.} Let $(\mathcal M, g)$ and $T_x^\ast \mathcal M$ denote the spacetime manifold and the cotangent space at $x \in \mathcal M$, respectively. The gas is described in terms of the one-particle distribution function $f \colon \Gamma_m^+ \to \mathbb R$ defined on the future mass shell $\Gamma_m^+ := \bigcup_{x \in \mathcal M} \{(x,p) \colon p \in P_x^+(m) \}$, where
\begin{eqnarray}
P_x^+(m) & := & \{ p \in T_x^\ast \mathcal{M} \colon g^{\mu\nu}(x) p_\mu p_\nu = - m^2, \nonumber \\
&& p \text{ is future directed} \}.
\end{eqnarray}
The distribution function satisfies the Vlasov equation
\begin{equation}
g^{\mu \nu} p_\nu \frac{\partial f}{\partial x^{\mu}}
 - \frac{1}{2}\frac{\partial g^{\alpha\beta}}{\partial x^{\mu}}p_{\alpha}p_{\beta}
 \frac{\partial f}{\partial p_{\mu}} = 0.
 \label{Eq:Vlasov}
\end{equation}
The particle current density associated with the distribution function $f$ can be computed as
\begin{equation}
J_\mu(x) := \int\limits_{P_x^+(m)} p_\mu f(x,p) \mbox{dvol}_x(p),
\label{Eq:Observables}
\end{equation}
where $\mathrm{dvol}_x(p)$ denotes the volume form on $P_x^+(m)$. Likewise, the energy-momentum-stress tensor is defined as
\begin{equation}
T_{\mu\nu}(x) := \int\limits_{P_x^+(m)} p_\mu p_\nu f(x,p) \mbox{dvol}_x(p).
\label{Eq:Tmunu}
\end{equation}
It is convenient to introduce the invariant particle number density defined by $n := \sqrt{-J_\mu J^\mu}$.

In horizon-penetrating coordinates $(t,r,\vartheta,\varphi)$ the Kerr metric can be written as
\begin{eqnarray}
g & = & -dt^2 +dr^2 -2a\sin^2\vartheta dr d\varphi 
 + \left( r^2 + a^2 \right)\sin^2 \vartheta d\varphi ^2  \nonumber \\
 && 
 + \rho^2 d\vartheta ^2 
 + \frac{2M r}{\rho^2}  \left(dt + dr - a\sin ^2 \vartheta d\varphi \right)^2,
\label{Eq:Kerr}
\end{eqnarray}
where $\rho^2 := r^2 + a^2\cos^2 \vartheta$, and $M$ and $a$ denote the black hole mass and the rotation parameter, respectively. We restrict ourselves to the sub-extremal case, such that the spin parameter $\alpha:=a/M$ lies in the interval $-1 < \alpha < 1$. The Kerr spacetime admits the two Killing vector fields $k = \partial_t$ and $\eta = \partial_\varphi$.

Geodesic equations are separable~\cite{bC68,mWrP70} in the coordinates $(t,r,\vartheta,\varphi)$ as follows: $p_t = - E$, $p_\varphi =L_z$, $p_\vartheta^2 = \Theta(\vartheta)$, $\left( \Delta p_r - 2M E r + a L_z \right)^2 = R(r)$, where $\Delta:=r^2 - 2Mr + a^2$,
\begin{subequations}
\begin{eqnarray}
    \Theta(\vartheta) & = &  L^2 - \left(\frac{L_z}{\sin\vartheta} - a\sin \vartheta E \right)^2 
 - a^2 m^2 \cos^2\vartheta, \\
    R(r) & = & \left[ E(r^2 + a^2 ) - a L _z  \right]^2 - \Delta\left(L^2 + m^2 r^2 \right),
\end{eqnarray}
\end{subequations}
and where $m$, $E$, $L_z$, and $L$ denote constants of motion.

In what follows, we specialize to distribution functions of the form $f(x,p) = F_\infty(E)$ for $(x,p) \in U$, and $f = 0$ outside of $U$, where $U$ denotes the invariant phase-space region which is relevant for the model. It is simple to verify that any $f$ of this type satisfies Eq.\ (\ref{Eq:Vlasov}). In terms of the momentum components $(p_t,p_r,p_\vartheta,p_\varphi)$, the volume form $\mathrm{dvol}_x(p)$ can be expressed as
\begin{equation}
    \mathrm{dvol}_x(p) = \sqrt{-\det(g^{\mu\nu})} \frac{dp_r dp_\vartheta dp_\varphi}{p^t}.
\end{equation}
In principle, this allows one to compute the components of the particle current density $J_\mu$, provided that the set $U$ is known. In the current model, we only take into account unbound orbits originating at infinity. They can either plunge into the black hole (we refer to those orbits, as well as to all corresponding quantities, as absorbed) or be scattered by the centrifugal barrier (they will be referred to as scattered ones). We neglect any contributions from bound orbits, as well as orbits emanating from the white hole.

Phase-space coordinates adapted to the description of unbound orbits are suggested by the polar equation $p_\vartheta^2 = \Theta(\vartheta)$. We substitute for $L$ and $L_z$ a new variable $Q > 0$ and an angle $\chi$, such that $Q^2 = L^2 - a^2 m^2 \cos^2\vartheta$ and
\begin{subequations}
\begin{eqnarray}
p_\vartheta &=& Q\cos\chi,\\
L_z &=& Q\sin\vartheta\sin\chi + a\sin^2\vartheta E.
\end{eqnarray}
\end{subequations}
The coordinates $(E,Q,\chi)$ were independently introduced in \cite{yL25} in the broader context of the Kerr-Newman spacetime. The proof that they cover the relevant phase-space region corresponding to unbound orbits is given in \cite{mainpaper}.

In terms of the variables $(E,Q,\chi)$, the volume form $\mbox{dvol}_x(p)$ becomes
\begin{equation}
\mbox{dvol}_x(p) = \frac{dE (Q dQ) d\chi}{\sqrt{R(r)}},
\end{equation}
where the radial potential $R(r)$ expressed in terms of $E$, $Q$, and $\chi$ reads
\begin{equation}
R(r) = \left( E\rho^2 - aQ\sin\vartheta\sin\chi \right)^2 - \Delta(Q^2 + m^2\rho^2).
\label{Eq:RQchi}
\end{equation}

Unbound orbits correspond to the following ranges of $E$, $Q$, and $\chi$: $E > m$, $0 < \chi < 2\pi$. Particles plunging into the black hole are characterized by $0 < Q < Q_c$, whereas the range $Q_c < Q < Q_\mathrm{max}$ corresponds to scattered orbits. Here, $Q_c$ is a smooth function of $a$, $\vartheta$, $\sin\chi$, and $E$, and it can be expressed in parametric form. The corresponding characteristic radius will be denoted by $r_c$ (see \cite{mainpaper}). The upper limit $Q_\mathrm{max}$ can be expressed explicitly as
\begin{widetext}
\begin{equation}
Q_\mathrm{max} = \frac{\rho(\rho^2 E^2 - \Delta m^2)}{\rho E a\sin\vartheta\sin\chi + \sqrt{\Delta}\sqrt{\rho^2 E^2 - \Delta m^2 + m^2 a^2\sin^2\vartheta\sin^2\chi}}.
\label{Eq:Qpm}
\end{equation}
\end{widetext}
A detailed proof of this characterization is involved and can be found in the accompanying paper \cite{mainpaper}. 


\begin{figure}
\begin{center}
\includegraphics[width=1.0\linewidth]{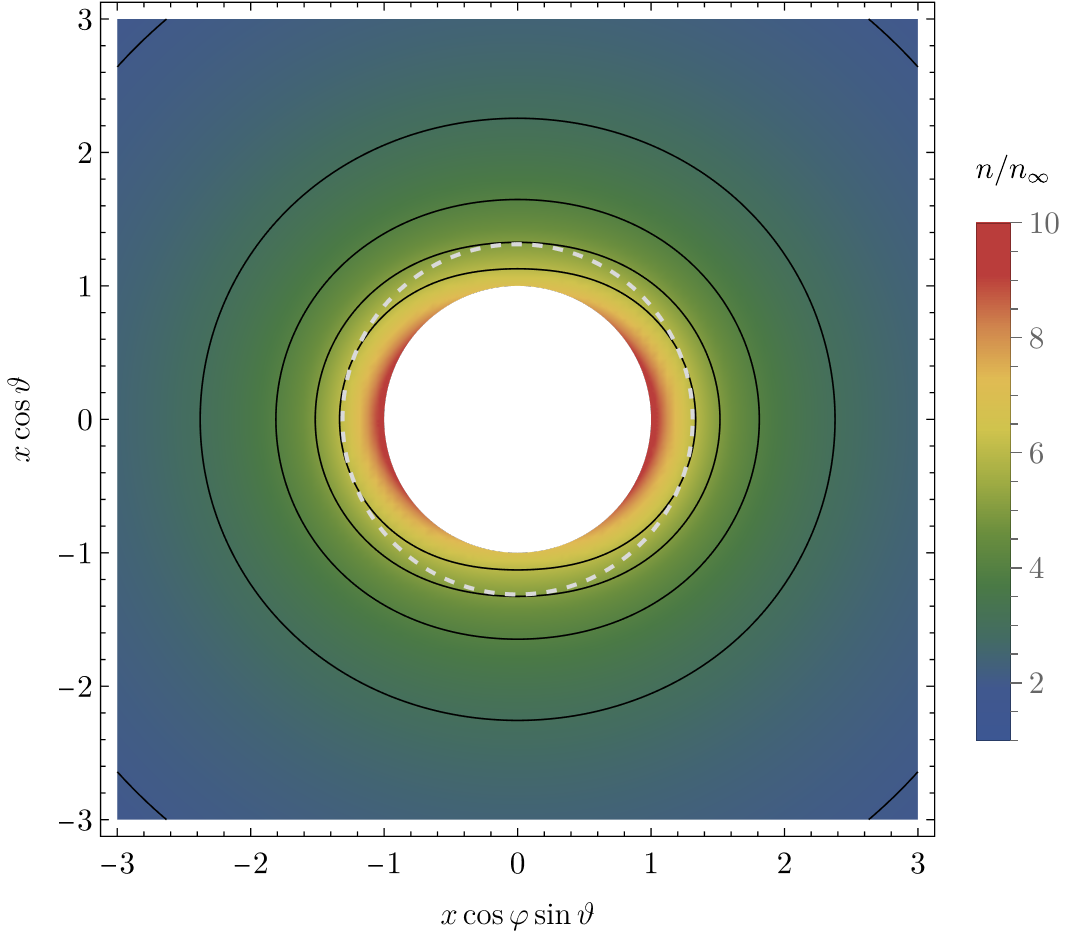}
\end{center}
\caption{\label{fig:compratio} Compression ratio at the meridional plane close to the black hole horizon (white dashed line) for the Maxwell-J\"{u}ttner model with $z = 1$ and $\alpha = 0.95$. We denote $x = r/M$. The solid dark lines mark contours of constant $n$. This plot can be compared with Fig.\ 11 in~\cite{mainpaper} obtained for monoenergetic distributions. }
\end{figure}

\textit{Bondi-type solution.} The components of the particle current density associated with absorbed and scattered orbits can be expressed as
\begin{subequations}
\begin{eqnarray}
J_t^\mathrm{(abs)} &=& -\int\limits_m^\infty dE E F_\infty(E) \int\limits_0^{2\pi} d\chi \int\limits_0^{Q_c} \frac{Q dQ}{\sqrt{R}},
\label{Eq:Jtabs}\\
J_t^\mathrm{(scat)} &=& -2\int\limits_m^\infty dE E F_\infty(E) \int\limits_0^{2\pi} d\chi 1_{r_c < r}
\int\limits_{Q_c}^{Q_\mathrm{max}} \frac{Q dQ}{\sqrt{R}}, 
\qquad
\label{Eq:Jtscat}
\end{eqnarray}
\end{subequations}
\begin{eqnarray}
J^r_\mathrm{(abs)} &=& - \frac{1}{\rho^2} \int\limits_m^\infty dE F_\infty(E) \int\limits_0^{2\pi} d \chi \int\limits_0^{Q_c} Q dQ,
\label{Eq:Jrabs}
\end{eqnarray}
and
\begin{subequations}
\begin{eqnarray}
J_\varphi^\mathrm{(abs)} &=& \sin \vartheta \left[ - a \sin \vartheta J_t^\mathrm{(abs)} \right. \nonumber \\
& & \left. + \int\limits_m^\infty dE F_\infty(E) \int\limits_0^{2\pi} d\chi \sin \chi \int\limits_0^{Q_c} \frac{Q^2 dQ}{\sqrt{R}} \right], 
\label{Eq:Jphiabs}\\
J_\varphi^\mathrm{(scat)} &=& \sin \vartheta \left[ - a \sin \vartheta J_t^\mathrm{(scat)} \right. \label{Eq:Jphiscat} \\
&& \left. + 2 \int\limits_m^\infty dE F_\infty(E) \int\limits_0^{2\pi} d\chi 1_{r_c < r}\sin\chi \int\limits_{Q_c}^{Q_\mathrm{max}} \frac{Q^2 dQ}{\sqrt{R}} \right]. \nonumber
\end{eqnarray}
\end{subequations}
The remaining components $J^r_\mathrm{(scat)}$, $J_\vartheta^\mathrm{(abs)}$, and $J_\vartheta^\mathrm{(scat)}$ vanish. Here, the characteristic function $1_{r_c < r}$ is equal to 1 if $r_c < r$ and 0, otherwise. 

In \cite{mainpaper} we consider distribution functions of monoenergetic particles $F_\infty = A m \delta(E - E_0)$ or an asymptotic Maxwell-J\"{u}ttner distribution $F_\infty = A \exp(-z E/m)$, where $z = m/(k_\mathrm{B} T)$. Here, $T$ denotes the associated asymptotic temperature and $k_\mathrm{B}$ is the Boltzmann constant. In this letter, we focus on Maxwell-J\"{u}ttner models. The corresponding asymptotic particle number density reads
\begin{equation}
n_\infty = 4 \pi m^3 A \frac{K_2(z)}{z},
\end{equation}
where $K_j(z)$ denotes the modified Bessel function of the second kind \cite{DLMF}. It can be used to replace the constant $A$, which is otherwise difficult to control in physical terms. Alternatively, one can use the asymptotic energy density defined as $\varepsilon_\infty = \lim\limits_{r\to\infty} T_{\mu\nu}k^\mu k^\nu$, which reads
\begin{equation}
\varepsilon_\infty = 4 \pi m^4 A \left[ \frac{K_1(z)}{z} + 3 \frac{K_2(z)}{z^2} \right].
\end{equation}

Figure~\ref{fig:compratio} shows the compression ratio $n/n_\infty$ in the meridional plane for a black hole with spin parameter $\alpha = 0.95$ and the inverse temperature $z = 1$. A polar-to-equatorial asymmetry is clearly visible, the compression ratio at the equator being higher than at the poles for points lying close to the horizon.


\begin{figure*}
\begin{center}
\includegraphics[width=0.49\textwidth]{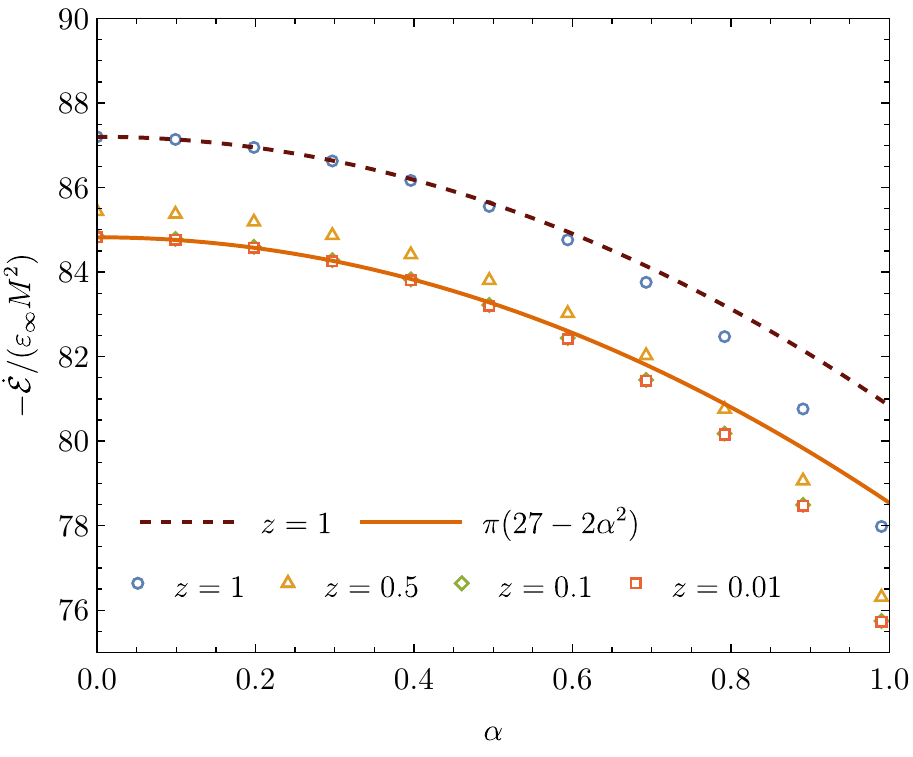}
\includegraphics[width=0.49\textwidth]{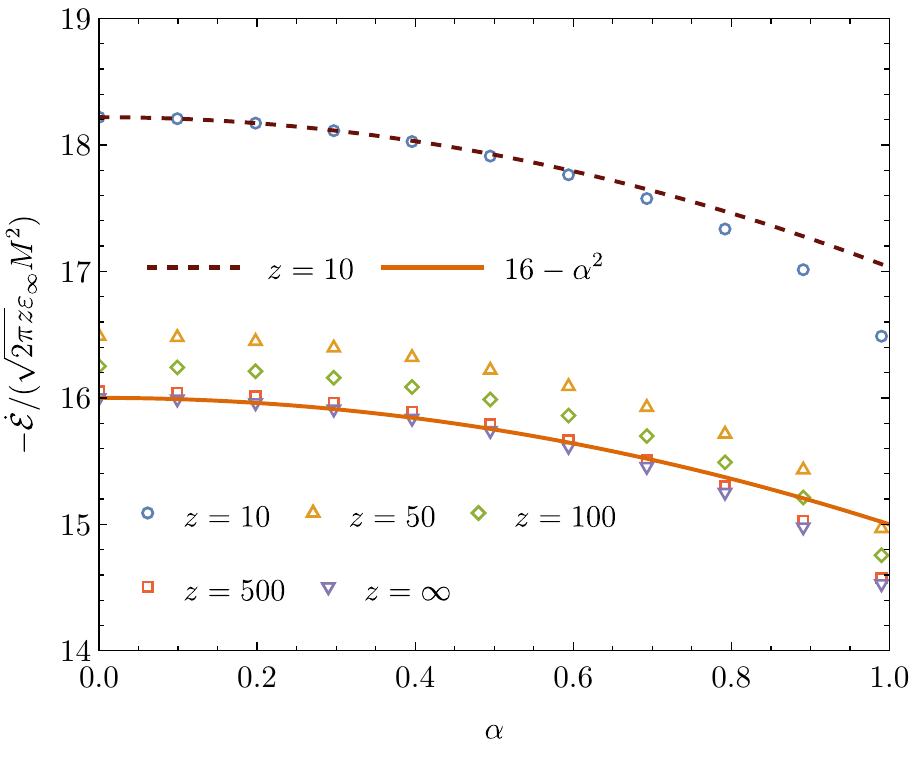}
\end{center}
\caption{\label{fig:accretionrates} Energy accretion rates for the Maxwell-J\"{u}ttner model with small (left) and large (right) values of $z$. Discrete marks denote exact values. Lines correspond to the slow-rotation approximation.
}
\end{figure*}

\textit{Accretion rates.} The three conserved currents $J^\mu$, $T\indices{^\mu_\nu} k^\nu$, and $T\indices{^\mu_\nu} \eta^\nu$ give rise to accretion rates corresponding to particle $\dot{\mathcal N}$, energy $\dot{\mathcal E}$, and angular momentum $\dot{\mathcal J}$, respectively. They can be expressed as
\begin{subequations}
\begin{eqnarray}
\mathcal{\dot N} &=& -\pi\int\limits_m^\infty dE F_\infty(E) \int\limits_0^\pi d\vartheta \sin\vartheta \int\limits_0^{2\pi} d\chi (Q_c)^2,
\label{Eq:dotN}\\
\mathcal{\dot E} &=& -\pi\int\limits_m^\infty dE E F_\infty(E) \int\limits_0^\pi d\vartheta \sin\vartheta \int\limits_0^{2\pi} d\chi (Q_c)^2,
\label{Eq:dotE}\\
\mathcal{\dot J} &=&  -2\pi \int\limits_m^\infty dE F_\infty(E) \int\limits_0^\pi d \vartheta \sin^2 \vartheta 
\nonumber\\
&& \times\int\limits_0^{2 \pi} d\chi \left[ \frac{1}{3} \sin \chi Q_c^3 + \frac{a}{2} \sin \vartheta E Q_c^2 \right].
\end{eqnarray}
\end{subequations}
The accretion rates $\dot {\mathcal N}$, $\dot {\mathcal E}$, and $\dot {\mathcal J}$ can be computed numerically or approximated in the slow-rotation limit. Including third-order contributions in $\alpha$, one obtains
\begin{widetext}
\begin{subequations}
\label{Eq:AccretionRateMJapp}
\begin{eqnarray}
\frac{\dot{\mathcal{N}}}{n_\infty}&=&-\frac{\pi z M^2}{K_2(z)}\int_1^\infty\frac{\left(x_c^{(0)}\right)^2}{x_c^{(0)}-3}\left[ 1 - \frac{2\alpha^2}{\left(x_c^{(0)}\right)^2\left(6 - x_c^{(0)}\right)} \right]e^{-z \varepsilon} d\varepsilon +  \mathcal{O}(\alpha^4),\\
\frac{\dot{\mathcal{E}}}{\varepsilon_\infty}&=&-\frac{\pi z^2  M^2}{zK_1(z)+3K_2(z)}\int_1^\infty\frac{\left(x_c^{(0)}\right)^2}{x_c^{(0)}-3}\left[ 1 - \frac{2\alpha^2}{\left(x_c^{(0)}\right)^2\left(6 - x_c^{(0)}\right)} \right] \varepsilon e^{-z \varepsilon} d\varepsilon + \mathcal{O}(\alpha^4),\\
\frac{\dot{\mathcal{J}}}{\varepsilon_\infty} &=& \frac{4 \pi \alpha z^2 M^3}{3[zK_1(z)+3K_2(z)]}\int_1^\infty  \left(\frac{x_c^{(0)}}{x_c^{(0)}-3} \right)^{\frac{3}{2}} \left\{1+ \frac{3 \alpha ^2 \left(x_c^{(0)}-3\right) \left[\left(x_c^{(0)}\right)^2-24\right]}{5 \left(x_c^{(0)}-6\right)^3 \left(x_c^{(0)}\right)^2} \right\} e^{-z \varepsilon} d\varepsilon + \mathcal{O}(\alpha^4),
\end{eqnarray}
\end{subequations}
\end{widetext}
where
\begin{equation}
x_c^{(0)} = \frac{8}{\varepsilon\sqrt{9\varepsilon^2 - 8} + 4 - 3\varepsilon^2}
\label{Eq:xc0}
\end{equation}
describes the zeroth-order (Schwarzschild) contribution to the critical radius $r_c = M x_c$ at given energy $E = m\varepsilon$.

An even simpler expression can be obtained in the low ($z \to \infty$) and high ($z \to 0$) asymptotic temperature limits. Neglecting terms that are  quartic or higher order in $\alpha$, one gets
\begin{subequations}
\label{Eq:Extreme-temperatureTE}
\begin{eqnarray}
\frac{\mathcal{\dot{N}}}{n_{\infty }} &=& \frac{\mathcal{\dot{E}}}{\varepsilon_\infty} 
 = -\mathcal{A}(1 - \mathcal{B}\alpha^2) M^2,
\\
\frac{\mathcal{\dot{J}}}{\varepsilon_{\infty }} &=& 2\delta\mathcal{A}(1 + \gamma\alpha^2)M^3\alpha,
\end{eqnarray}
\end{subequations}
where
\begin{eqnarray}
\mathcal A &=& 16 \sqrt{2 \pi z}, \quad \mathcal B=\frac{1}{16}, \quad \delta=\frac{1}{3}, \quad \gamma=\frac{3}{80}, \quad z \to \infty,
\nonumber\\
\mathcal A &=& 27\pi, \quad \mathcal B=\frac{2}{27}, \quad \delta=\frac{2}{3}, \quad \gamma=0, \quad z \to 0.
\nonumber
\end{eqnarray}
The exact values of the energy accretion rates are illustrated in Fig.\ \ref{fig:accretionrates}, both for small and large values of $z$. They depend weakly on the black hole spin parameter and hence remain in excellent agreement with the results of the slow-rotation approximation, beyond the expected range of $\alpha$. We checked that for several values of $z$ in the range $z\in(0,\infty)$, the difference in the energy and angular momentum accretion rates between the results obtained within the slow-rotation approximation and the exact numerical values is smaller than $4\%$ for the energy accretion rate and below $5\%$ for angular momentum accretion rate for black holes with $\alpha \leq 0.99$.


\textit{Discussion.} 
In the quasistationary approximation, the changes of the black hole mass $M$ and the angular momentum $J = aM = \alpha M^2$ can be estimated from $\dot{\mathcal E}$ and $\dot{\mathcal J}$. In the cosmological context this approximation was already used in \cite{yZiN67}. To some extent, the quasistationary approximation was justified by a recent numerical study \cite{chRtBsS21}. The fact that $-\dot{\mathcal E}$ should be associated with the rate of the black hole mass growth can be shown for self-gravitating perfect fluids in spherical symmetry \cite{pMeM22}; see also \cite{dC70} where the corresponding assumption was made for individual test particles plunging into the black hole. In our case, Eqs.~(\ref{Eq:Extreme-temperatureTE}) yield
\begin{subequations}
\label{Eq:MdotJdot}
\begin{eqnarray}
\frac{d M}{dt} &=& - \dot{\mathcal E}=\varepsilon_\infty \mathcal A (1-\mathcal{B} \alpha^2) M^2,\label{Eq:Mdot}\\
\frac{dJ}{dt} &=& - \dot{ \mathcal J}=-2 \delta \varepsilon_\infty \mathcal{A} (1+\gamma \alpha^2) M^3 \alpha,
\label{Eq:Jdot}
\end{eqnarray}
which, using the relation $J = \alpha M^2$, implies a decrease of the spin parameter's magnitude at the rate
\begin{equation}
\frac{d\alpha}{dt} =-2 \varepsilon_\infty \mathcal{A} (1+\delta ) (1-\mathcal C \alpha^2) M \alpha, \quad \mathcal C = \frac{\mathcal B-\delta \gamma}{1+\delta}.
\label{Eq:Adot}
\end{equation}
\end{subequations}
Equations~(\ref{Eq:Mdot}) and (\ref{Eq:Adot}) govern the time evolution of the black hole mass and spin parameter, and they yield the following relation:
\begin{equation}
\frac{M_0}{M}=\left( \frac{\alpha}{\alpha_0} \right)^\frac{1}{2+2\delta} \left( \frac{1-\mathcal C \alpha^2}{1-\mathcal C \alpha_0^2} \right)^{\frac{\mathcal B- \mathcal C}{4 (1+\delta) \mathcal C}},
\end{equation}
where $\alpha_0$ and $M_0$ denote, respectively, the spin parameter and the black hole mass at the initial time $t_0$. Substituting the resulting expression for the mass $M$ in Eq.~(\ref{Eq:Adot}), one obtains an equation for $\alpha$ which, in principle, can be integrated.

We conclude this letter with two applications of Eqs.~(\ref{Eq:MdotJdot}). For these applications, it will turn out to be sufficient to neglect the relatively small term $\mathcal B \alpha^2$ in Eq.~(\ref{Eq:Mdot}) which decreases during the evolution. Hence, restoring SI units and neglecting terms that are quadratic or higher order in the rotation parameter, we can integrate Eq.~(\ref{Eq:Mdot}) from $t_0$ to $t_X$, corresponding to a mass growth from $M_0$ to $M = X M_0$ and a decrease of the spin parameter from $\alpha_0$ to $\alpha=\alpha_0/X^{2+2\delta}$,  as
\begin{equation}
1-\frac{1}{X}=\frac{G^2}{c^5} \mathcal A M_0 \int_{t_0}^{t_X} \varepsilon_\infty(\bar t) d \bar t,
\label{Eq:AccretionRate}
\end{equation}
where we have allowed for the energy density to be time-dependent. The first application considers a cosmological scenario consisting of the growth of primordial black holes, due to the
accretion of hot dark matter particles, following~\cite{yZiN67}
and~\cite{pMoA21b}. In this case, $\mathcal{A}=27\pi$ is constant and $\varepsilon_\infty = 3c^2/(32\pi G t^2)$. It then follows from Eq.~(\ref{Eq:AccretionRate}) that the mass diverges and the black hole spin parameter $\alpha$ tends to zero in a finite time $t_\infty$, given by
\begin{equation}
\frac{t_0}{t_\infty} = 1 - \frac{32}{81\kappa},
\end{equation}
where $\kappa$ is a dimensionless parameter of the order of unity defined by $M_0 = \kappa c^3 t_0/G$, as in~\cite{pMoA21b}, and it was assumed that $81\kappa > 32$. Therefore, one is led to the same conclusion as in the Schwarzschild model: if $\kappa\sim 1$ the accretion process by primordial black holes is robust, even if they are highly spinning initially.

The second scenario corresponds to a supermassive black hole which accretes dark matter from its environment.  Assuming a reservoir of constant density and low temperature, neglecting contributions from baryonic matter which does not interact directly with the dark matter, and integrating Eq.~(\ref{Eq:AccretionRate}), one finds that the time required for the black hole mass to grow by a factor $X$ (and the black hole spin parameter to decrease from $\alpha_0$
to $\alpha=\alpha_0/X^{2+2\delta}$) is $t_X = (1-1/X) c^5/(G^2 M_0\mathcal A \varepsilon_\infty)$, where for simplicity we have set $t_0 = 0$. Therefore, the characteristic time scale $t_\infty$ for having a significant mass growth is related to the parameter $z$ according to
\begin{equation}
\sqrt{z} = \frac{c^5}{\sqrt{512\pi} G^2 M_0 \varepsilon_\infty  t_\infty},
\end{equation}
which leads to
\begin{equation}
\frac{k_\mathrm{B} T}{m c^2} \sim
0.5\times 10^{-18}
\left( \frac{M_0}{10^9 M_\odot} \right)^2
\left( \frac{\varepsilon_\infty}{1\frac{\mathrm{GeV}}{\mathrm{cm^3}}} \right)^2
\left( \frac{t_\infty}{10^9 \, \mathrm{yr}} \right)^2.
\end{equation}
Taking as an example $t_\infty = 10^9$ yr and the supermassive black hole in the center of M87, for which $M_0 \simeq 6.5\times 10^9 M_\odot$~\cite{EHTVI} and the density of dark matter within the core is $\varepsilon_\infty\simeq 5\times 10^{-25} \, \mathrm{g/cm^3} \times c^2\simeq 0.3 \,\mathrm{GeV/cm^3}$~\cite{dLpS22}, one obtains the order-of-magnitude estimate $k_\mathrm{B} T/(m c^2)\sim 2\times 10^{-18}$. This shows that dark matter needs to be extremely cold in order to lead to a significant mass growth of the black hole as in the scenario considered in~\cite{dLpS22}. Furthermore, notice that our estimate lies below the cosmological upper bound of $10^{-14}$ in~\cite{cAjN14}.

\textit{Acknowledgments.} We thank Emilio Tejeda for fruitful discussions. P.\ M.\ acknowledges a support of the Polish National Science Centre Grant No.\ 2017/26/A/ST2/00530. M.\ M.\ was supported by SECIHTI through Estancias
Posdoctorales por M\'exico Convocatoria 2023(1) under the postdoctoral Grant No.~1242413. O.\ S.\ was partially supported by CIC Grant No.~18315 to Universidad Michoacana and by CONAHCyT Network Project No.~376127 ``Sombras, lentes y ondas gravitatorias generadas por objetos compactos astrof\'isicos". Finally, M.\ M.\ and O.\ S.\ acknowledge financial support from SECIHTI-SNII. 

\textit{Data availability.} The data that support the findings of this article are openly available \cite{articleDataFile2026}.

\bibliographystyle{unsrt}
\bibliography{refs_kinetic}

\end{document}